\documentclass{article}
\usepackage[a4paper,inner=30mm,outer=30mm,top=20mm,bottom=40mm,includehead]{geometry}
\usepackage{amsmath}
\numberwithin{equation}{section}

\begin{document}
\thispagestyle{empty}
\title{\vspace{-2cm}
\bf A concise introduction to trapped surface formation in general relativity}

\author{Donato Bini, Istituto M. Picone, CNR, Rome, Italy \\
Giampiero Esposito, INFN Sezione di Napoli, Naples, Italy}

\date{}
\maketitle

\begin{abstract}
Trapped surface formation in general relativity can be studied through a coupled set of
nonlinear equations, where various terms can be neglected, as 
was proved by a rigorous mathematical analysis of Christodoulou.
This paper is devoted to a pedagogical synthesis of the mathematical formalism 
employed in this analysis, i.e. the optical structure of general relativity 
investigated by Christodoulou, Klainerman and other authors after them.
\end{abstract}

\section{Introduction}
\setcounter{equation}{0}

In the sixties, when the global techniques of differential topology were applied to investigate 
causal structure and singularities in gravitational collapse and cosmology, the concept of
trapped surface was elaborated for the first time. In Ref. \cite{PE1965}, a trapped surface $\Sigma$
is defined to be a compact spacelike two-surface with the property that the null geodesics which
meet $\Sigma$ orthogonally, locally converge in future directions.
More precisely, this means what follows:
we consider in a 4-dimensional spacetime $(V,g)$ the compact spacelike 2-surface $\Sigma$ 
(i.e. a compact Riemannian 2-manifold) which is a submanifold of 
a 3-dimensional spacelike submanifold $M$.
Let $n$ and $\nu$ be the future unit normal to $M$  and $\Sigma$, respectively, with $\nu$ tangent to $M$. 
One can then define two null vectors
$\ell^\pm \equiv n \pm \nu$ which contribute to form the spacetime tensor 
\begin{equation}
h=g+n\otimes n - \nu \otimes \nu =g +\frac{1}{2}(\ell^+ \otimes \ell^- +\ell^- \otimes \ell^-).
\label{(1.1)}
\end{equation}
These tensors are then used to define the null mean curvatures
(in our paper we always use explicit summations to achieve a clear distinction
between 2-, 3- and 4-dimensional concepts)
\begin{equation}
\chi^{\pm} \equiv \sum_{\alpha,\beta=1}^{4}
h^{\alpha \beta}\nabla_{\alpha} \ell^\pm_{\beta}
= \sum_{\alpha=1}^{4} \nabla_\alpha \ell^\pm{}^\alpha.
\label{(1.2)}
\end{equation}
The surface $\Sigma$ is said to be a trapped surface in the spacetime $(V,g)$ 
if both $\chi^{+}$ and $\chi^{-}$ are negative on $\Sigma$, i.e., 
if the null geodesic congruences both converge \cite{Choquet-Bruhat:2009xil,Senovilla,Beng}.

Penrose asked himself the question
whether it is reasonable to expect trapped surfaces to develop at all in our actual universe, and he
found that there can be no reason of principle against a trapped surface developing.
However, the proof of {\it theorems} on the formation of trapped surfaces for solutions of vacuum
Einstein equations was a much harder task, and was accomplished only in 2009 by Christodoulou
\cite{BH2009}, with an appropriate choice of initial conditions, called the short-pulse method. Later on,
a simpler proof of trapped-surface formation was obtained in Ref. \cite{KR2012} by enlarging the admissible 
set of initial conditions and relaxing the corresponding propagation estimates. 

The basic tool in such investigations is the geometry of double null foliations, which leads eventually
to the so-called optical structure of vacuum Einstein equations. Although the physics-oriented 
community is by now familiar with Newman-Penrose formalism and solutions of the eikonal equation in
curved spacetime, the techniques used in Refs. \cite{BH2009,KR2012} are not yet widespread, and hence we 
find it appropriate to summarize here some key concepts and results in such References. We therefore
consider a region ${\cal D}={\cal D}(u_{*},v_{*})$ of a vacuum spacetime $(M,g)$ spanned by a double null
foliation generated by the optical functions $(u,v)$ that solve the eikonal equation
\begin{equation}
({\rm grad}\varphi,{\rm grad}\varphi)=\sum_{\mu,\nu=1}^{4}(g^{-1})^{\mu \nu}
(\partial_{\mu}\varphi)(\partial_{\nu}\varphi)=0, \; \varphi=u,v
\label{(1.3)}
\end{equation}
and are increasing towards the future, so that $u$ takes values in the closed interval
$[0,u_{*}]$ and $v$ takes values in the closed interval $[0,v_{*}]$. Let $H_{u}$ be the outgoing
null hypersurfaces generated by the level surfaces of $u$, and let $H_{v}$ be the incoming null
hypersurfaces generated by the level surfaces of $v$. The two-dimensional surfaces 
obtained by intersection of $H_{u}$ and $H_{v}$ are
\begin{equation}
S_{uv} \equiv H_{u} \cap H_{v},
\label{(1.4)}
\end{equation}
and we denote by $H_{u}(v_{1},v_{2})$ the portion of $H_{u}$ defined by $v \in [v_{1},v_{2}]$,
while $H_{v}(u_{1},u_{2})$ is the portion of $H_{v}$ defined by $u \in [u_{1},u_{2}]$. The two
optical functions $u$ and $v$ make it possible to define the vector fields with components
given by
\begin{equation}
L^{\mu} \equiv -2 \sum_{\nu=1}^{4}(g^{-1})^{\mu \nu}\partial_{\nu}u,
\label{(1.5)}
\end{equation}
\begin{equation}
M^{\rho} \equiv -2 \sum_{\lambda=1}^{4}(g^{-1})^{\rho \lambda}\partial_{\lambda}v.
\label{(1.6)}
\end{equation}
These are future-directed null geodesic vector fields, in that
\begin{equation}
\nabla_{L}L=0, \; \nabla_{M}M=0,
\label{(1.7)}
\end{equation}
the integral curves of $L$ being the generators of each $H_{u}$, and the integral curves of $M$
being the generators of each $H_{v}$. Indeed, one has in arbitrary local coordinates
\cite{BH2009}
\begin{eqnarray}
\; & \; & \sum_{\mu,\nu=1}^{4}g_{\lambda \mu} L^{\nu}\nabla_{\nu}L^{\mu}
=-2 \sum_{\nu=1}^{4}L^{\nu}\nabla_{\nu}\partial_{\lambda}u
=-2\sum_{\nu=1}^{4}L^{\nu}\nabla_{\lambda}\partial_{\nu}u \nonumber \\
&=& 4 \sum_{\nu,\rho=1}^{4}(g^{-1})^{\nu \rho}(\partial_{\rho}u)
\nabla_{\lambda}\partial_{\nu}u
=2 \partial_{\lambda} \left(\sum_{\nu,\rho=1}^{4}(g^{-1})^{\nu \rho}
(\partial_{\nu}u)(\partial_{\rho}u)\right)=0,
\label{(1.8)}
\end{eqnarray}
where we have exploited the commutation of covariant derivatives of functions when torsion vanishes,
the condition $\nabla g=0$, the Leibniz rule to express
\begin{equation}
\sum_{\nu,\rho=1}^{4}(g^{-1})^{\nu \rho}\Bigr[(\partial_{\nu}u)(\nabla_{\lambda}\partial_{\rho}u)
+(\partial_{\rho}u)(\nabla_{\lambda}\partial_{\nu}u)\Bigr]
=\nabla_{\lambda}({\rm grad}u,{\rm grad}u)
\label{(1.9)}
\end{equation}
and the eikonal equation (1.1) for $\varphi=u$.
The same holds with $u$ and $L$ replaced by $v$ and $M$, respectively.

Once the geodesic vector fields $L$ and $M$ are at our disposal, we can define
\begin{equation}
{1 \over 2}\Omega^{2} \equiv -{1 \over g(L,M)}
=-{1 \over \sum_{\mu \nu=1}^{4}g_{\mu \nu}L^{\mu}M^{\nu}}.
\label{(1.10)}
\end{equation}
For small values of $u_{*}$ and $v_{*}$, the spacetime slab 
${\cal D}(u_{*},v_{*})$ is completely determined by data along the null characteristic
hypersurfaces $H_{0}$ and ${\widetilde H}_{0}$ corresponding to $v=0$ and $u=0$, respectively. 
We assume that $H_{0}$ can be extended \cite{KR2012} to negative values of $v$, and that the
spacetime $(M,g)$ is Minkowskian for $v<0$ and for all $u \geq 0$. The double null foliation
can be chosen in such a way that the function $\Omega$ defined in (1.8) obeys the condition
\begin{equation}
\Omega(0,v)=1 \; \; \forall v \in [0,v_{*}],
\label{(1.11)}
\end{equation}
and one defines the pair $(e_{3},e_{4})$ of null vector fields such that
\begin{equation}
e_{3} \equiv \Omega M, \; e_{4} \equiv \Omega L,
\label{(1.12)}
\end{equation}
\begin{equation}
g(e_{3},e_{4})=\Omega^{2}g(L,M)=-2.
\label{(1.13)}
\end{equation}
Given a two-surface $S(u,v)$ and an arbitrary frame $(e_{1},e_{2})$ tangent to it,
one can define the Ricci coefficients
\begin{equation}
\Gamma_{(\lambda)(\mu)(\nu)}=g \Bigr(e_{(\lambda)},D_{e_{(\nu)}}e_{(\mu)}\Bigr), 
\; \lambda,\mu,\nu=1,2,3,4.
\label{(1.14)}
\end{equation}
With the understanding that lower-case indices $a,b$ take only the values $1,2$, and that
for covariant derivatives with a subscript $D_{3} \equiv D_{e_{3}}, \; 
D_{4} \equiv D_{e_{4}}$,  the Ricci coefficients are completely 
determined by the components
\begin{equation}
\chi_{ab} \equiv g(D_{a}e_{4},e_{b}), 
\; {\widetilde \chi}_{ab} \equiv g(D_{a}e_{3},e_{b}),
\label{(1.15)}
\end{equation}
\begin{equation}
\eta_{a} \equiv -{1 \over 2}g(D_{3}e_{a},e_{4}),
\; {\widetilde \eta}_{a} \equiv -{1 \over 2}g(D_{4}e_{a},e_{3}),
\label{(1.16)}
\end{equation}
\begin{equation}
\omega \equiv -{1 \over 4}g(D_{4}e_{3},e_{4}),
\; {\widetilde \omega} \equiv -{1 \over 4}g(D_{3}e_{4},e_{3}),
\label{(1.17)}
\end{equation}
\begin{equation}
\zeta_{a} \equiv {1 \over 2}g(D_{a}e_{4},e_{3}) .
\label{(1.18)}
\end{equation}
For example, one has
\begin{equation}
\chi_{ab}=\sum_{\rho,\sigma,\beta=1}^{4}g_{\rho \sigma}
(\Omega L^{\rho})_{;\beta}(e_{a})^{\beta}(e_{b})^{\sigma}.
\label{(1.19)}
\end{equation} 

In order to display the null structure equations, we have to consider also the trace-free parts
of $\chi_{ab}$ and ${\widetilde \chi}_{ab}$, here denoted by 
${\widehat \psi}_{ab}(e_{4})$ and ${\widehat \psi}_{ab}(e_{3})$, where we exploit the possibility
to define, for any vector field $X$,
\begin{equation}
\psi_{ab}(X) \equiv g(D_{a}X,e_{b}),
\label{(1.20)}
\end{equation}
so that
\begin{equation}
\chi_{ab}=\psi_{ab}(e_{4}), \; 
{\widetilde \chi}_{ab}=\psi_{ab}(e_{3}).
\label{(1.21)}
\end{equation}
Moreover, we need $\nabla$, the induced covariant derivative operator on the surface $S(u,v)$,
the projection $\nabla_{3}$ (respectively $\nabla_{4}$) to $S(u,v)$ of the covariant derivative
$D_{3}$ with respect to the vector field $e_{3}$ (respectively $D_{4}$ with respect to $e_{4}$),
and the null curvature components
\begin{equation}
\alpha_{ab} \equiv R(e_{a},e_{4},e_{b},e_{4}), \;
{\widetilde \alpha}_{ab} \equiv R(e_{a},e_{3},e_{b},e_{3}),
\label{(1.22)}
\end{equation}
\begin{equation}
\beta_{a} \equiv {1 \over 2} R(e_{a},e_{4},e_{3},e_{4}), \;
{\widetilde \beta}_{a} \equiv {1 \over 2} R(e_{a},e_{3},e_{3},e_{4}),
\label{(1.23)}
\end{equation}
\begin{equation}
\rho \equiv {1 \over 4} R(Le_{4},e_{3},e_{4},e_{3}), \;
\sigma \equiv {1 \over 4} { }^{*}R(e_{4},e_{3},e_{4},e_{3}),
\label{(1.24)}
\end{equation}
where ${ }^{*}R$ is the Hodge dual of $R$. One then arrives at the null structure equations
\cite{KR2012}
\begin{equation}
\nabla_{4}\chi_{ab}=-\sum_{c=1}^{2} \chi_{ac} \chi_{\; b}^{c}-2 \omega \chi_{ab}-\alpha_{ab},
\label{(1.25)}
\end{equation}
\begin{equation}
\nabla_{3}{\widetilde \chi}_{ab}=-\sum_{c=1}^{2}{\widetilde \chi}_{ac} {\widetilde \chi}_{\; b}^{c}
-2 {\widetilde \omega} {\widetilde \chi}_{ab}-{\widetilde \alpha}_{ab},
\label{(1.26)}
\end{equation}
\begin{equation}
\nabla_{4}\eta_{a}=-\sum_{c=1}^{2}\chi_{a}^{\; c} (\eta_{c}-{\widetilde \eta}_{c})-\beta_{a},
\label{(1.27)}
\end{equation}
\begin{equation}
\nabla_{3}{\widetilde \eta}_{a}=\sum_{c=1}^{2}{\widetilde \chi}_{a}^{\; c} 
(\eta_{c}-{\widetilde \eta}_{c})+{\widetilde \beta}_{a},
\label{(1.28)}
\end{equation}
\begin{equation}
\nabla_{4}{\widetilde \omega}=2 \omega {\widetilde \omega}
+{3 \over 4}(\eta_{a}-{\widetilde \eta}_{a})(\eta^{a}-{\widetilde \eta}^{a})
-{1 \over 4}(\eta_{a}-{\widetilde \eta}_{a})(\eta^{a}+{\widetilde \eta}^{a})
-{1 \over 8}(\eta_{a}+{\widetilde \eta}_{a})(\eta^{a}+{\widetilde \eta}^{a})
+{1 \over 2}\rho,
\label{(1.29)}
\end{equation}
\begin{equation}
\nabla_{3}\omega=2 \omega {\widetilde \omega}
+{3 \over 4}(\eta_{a}-{\widetilde \eta}_{a})(\eta^{a}-{\widetilde \eta}^{a})
+{1 \over 4}(\eta_{a}-{\widetilde \eta}_{a})(\eta^{a}+{\widetilde \eta}^{a})
-{1 \over 8}(\eta_{a}+{\widetilde \eta}_{a})(\eta^{a}+{\widetilde \eta}^{a})
+{1 \over 2}\rho,
\label{(1.30)}
\end{equation}
supplemented by the constraint equations
\begin{equation}
{\rm div}{\widehat \psi}(e_{4})={1 \over 2} \nabla {\rm tr} \chi
-{1 \over 2}(\eta-{\widetilde \eta})
\left({\widehat \psi}(e_{4})-{1 \over 2}{\rm tr}\chi \right)-\beta,
\label{(1.31)}
\end{equation}
\begin{equation}
{\rm div}{\widehat \psi}(e_{3})={1 \over 2} \nabla {\rm tr} {\widetilde \chi}
+{1 \over 2}(\eta-{\widetilde \eta})
\left({\widehat \psi}(e_{3})-{1 \over 2}{\rm tr}{\widetilde \chi}\right)
+{\widetilde \beta},
\label{(1.32)}
\end{equation}
\begin{equation}
{\rm curl} \; \eta=-{\rm curl}{\widetilde \eta}
=\sigma+{\widehat \psi}(e_{3}) \wedge {\widehat \psi}(e_{4}),
\label{(1.33)}
\end{equation}
\begin{equation}
K=-\rho+{1 \over 2}{\widehat \psi}(e_{4}){\widehat \psi}(e_{3})
-{1 \over 4}({\rm tr} \; \chi)({\rm tr} \; {\widetilde \chi}),
\label{(1.34)}
\end{equation}
where $K$ is the Gauss curvature of the 2-surface $S$. From these equations one gets
in particular a pair of equations which play a key role in the formation of trapped
surfaces, i.e. \cite{KR2012}
\begin{equation}
\nabla_{4}{\rm tr} \; \chi+{1 \over 2}({\rm tr} \; \chi)^{2}
=-\sum_{a,b=1}^{2}{\widehat \psi}_{ab}{\widehat \psi}^{ab}
-2 \omega {\rm tr} \; \chi,
\label{(1.35)}
\end{equation}
\begin{equation}
\nabla_{3}{\widehat \psi}+{1 \over 2}({\rm tr} \; {\widetilde \chi}){\widehat \psi}
=\nabla {\widehat \otimes} \eta + 2 {\widetilde \omega}{\widehat \psi}(e_{4})
-{1 \over 2}({\rm tr} \; \chi){\widehat \psi}(e_{3})
+\eta {\widehat \otimes}\eta \equiv F,
\label{(1.36)}
\end{equation}
where the tensor products in Eq. (1.34) have components obtainable from the general
formula for pairs of $1$-forms $C=\sum_{a=1}^{2}C_{a}dx^{a}$ and
$E=\sum_{b=1}^{2}E_{b}dx^{b}$ on the $2$-surface $S$ \cite{BH2009}
\begin{equation}
(C {\widehat \otimes} E)_{ab}=C_{a}E_{b}+C_{b}E_{a}-g_{ab}\sum_{f=1}^{2}C_{f}E^{f}.
\label{(1.37)}
\end{equation}
Note that the above expression can be written also as  a symmetric-tracefree part (modulo a factor of 2)
\begin{equation}
C {\widehat \otimes} E= 2[C\otimes E]^{\rm STF}.
\label{(1.38)}
\end{equation}

\section{Approximate form of nonlinear equation}
\setcounter{equation}{0}

Following Ref. \cite{KR2012}, it is instructive to outline an approximate treatment of the
nonlinear equation responsible for trapped-surface formation. For this purpose, we assume that
spacetime is Minkowskian for $v<0$ and all non-negative values of $u$. The values of $v$ are
restricted to the closed interval $[0,\delta]$, where $\delta$ is positive and small. The radius of
the $2$-surface $S=S(u,v)$ is denoted by $r=r(u,v)$, i.e. $|S(u,v)|=4\pi r^{2}$, and
$r(0,0) \equiv r_{0}$. Further assumptions are as follows.
\vskip 0.3cm
\noindent
(i) For small values of $\delta$, $u$ and $v$ approach their flat-space values
$u \approx {1 \over 2}(t-r+r_{0})$ and $v \approx {1 \over 2} (t+r-r_{0})$, while
$\Omega \approx 1$ and ${dr \over du} \approx -1$.
\vskip 0.3cm
\noindent
(ii) The value of ${\rm tr}{\widetilde \chi}$ is close to $-{2 \over r}$, corresponding to the
imbedding in flat space.
\vskip 0.3cm
\noindent
(iii) The right-hand side $F$ of Eq. (1.36) can be neglected in a first approximation, as well as
$-2 \omega {\rm tr}\chi$ on the right-hand side of Eq. (1.35).

In light of these assumptions, Eq. (1.35) reduces to
\begin{equation}
{d \over dv}{\rm tr}\chi \leq -|{\widehat \psi}|^{2},
\label{(2.1)}
\end{equation}
which, by integration, yields
\begin{equation}
{\rm tr}\chi(u,v) \leq {\rm tr} \chi(u,0)-\int_{0}^{v}|{\widehat \psi}|^{2}(u,v)dv
={2 \over r(u,0)}-\int_{0}^{v}|{\widehat \psi}|^{2}(u,v)dv.
\label{(2.2)}
\end{equation}
Now we can multiply the exact form of Eq. (1.36) by $\widehat \psi$, finding
\begin{equation}
{d \over du}|{\widehat \psi}|^{2}+({\rm tr}{\widetilde \chi})|{\widehat \psi}|^{2}
={\widehat \psi}F,
\label{(2.3)}
\end{equation}
while, by application of the Leibniz rule, adding and subtracting terms that make it possible
to exploit the assumptions (i) and (ii), we find
\begin{eqnarray}
{d \over du}(r^{2}|{\widehat \psi}|^{2})&=&r^{2}{d \over du}|{\widehat \psi}|^{2}
+2r {dr \over du}|{\widehat \psi}|^{2}
=r^{2}|{\widehat \psi}|^{2}\left(-{\rm tr}{\widetilde \chi}
+{2 \over r}{dr \over du}\right)+r^{2}{\widehat \psi}F 
\nonumber \\
&=&r^{2}|{\widehat \psi}|^{2}\left[-\left({\rm tr}{\widetilde \chi}
+{2 \over r}\right)+{2 \over r}\left(1+{dr \over du}\right)\right]
+r^{2}{\widehat \psi}F \equiv {\cal F},
\label{(2.4)}
\end{eqnarray}
which yields, upon integration,
\begin{equation}
r^{2}|{\widehat \psi}|^{2}(u,v)=r^{2}(0,v)|{\widehat \psi}|^{2}(0,v)
+\int_{0}^{u}{\cal F}(u',v)du'.
\label{(2.5)}
\end{equation}
By virtue of the assumptions (i) and (ii) and of Eq. (2.4), the integral 
$\int_{0}^{u}{\cal F}(u',v)du'$ is negligible in the slab ${\cal D}(u,\delta)$, and hence
one obtains the approximate relation
\begin{equation}
r^{2}|{\widehat \psi}|^{2}(u,v) \approx r^{2}(0,v)
|{\widehat \psi}|^{2}(0,v).
\label{(2.6)}
\end{equation}
As a next step, one freely prescribes the trace-free part of the extrinsic curvature
along the initial hypersurface \cite{KR2012}, so that
\begin{equation}
{\widehat \psi}(0,v)={\widehat \psi}_{0}(v)
\label{(2.7)}
\end{equation}
for some traceless $2$-tensor ${\widehat \psi}_{0}$. Hence Eq. (2.6) becomes
\begin{equation}
|{\widehat \psi}|^{2}(u,v) \approx {r^{2}(0,v) \over r^{2}(u,v)}
|{\widehat \psi}_{0}|^{2}(v).
\label{(2.8)}
\end{equation}
Furthermore, since $|v| \leq \delta$ and $r(u,v)=r_{0}+v-u$, Eq. (2.8) reduces to
\begin{equation}
|{\widehat \psi}|^{2}(u,v) \approx {r_{0}^{2} \over (r_{0}-u)^{2}}
|{\widehat \psi}_{0}|^{2}(v).
\label{(2.9)}
\end{equation}
This formula can be now inserted into the right-hand side of Eq. (2.2), and leads to
\begin{equation}
{\rm tr}\chi(u,v) \leq {2 \over (r_{0}-u)}
-{r_{0}^{2}\over (r_{0}-u)^{2}}
\int_{0}^{v}|{\widehat \psi}_{0}|^{2}(v')dv'
+{\rm error} \; {\rm term}. 
\label{(2.10)}
\end{equation}
It is now clear that the trace of the extrinsic curvature is never positive provided that
\begin{equation}
{2(r_{0}-u)\over r_{0}^{2}} < \int_{0}^{\delta}|{\widehat \psi}_{0}|^{2}(v')dv'.
\label{(2.11)}
\end{equation}
On the other hand, from Eq. (1.35) the condition for the initial hypersurface not to
contain trapped surfaces is
\begin{equation}
\int_{0}^{\delta}|{\widehat \psi}_{0}|^{2}(v')dv' < {2 \over r_{0}}.
\label{(2.12)}
\end{equation}
The joint effect of majorizations (2.11) and (2.12) is that formation of trapped surfaces
is expected provided that the condition
\begin{equation}
{2(r_{0}-u)\over r_{0}^{2}} < \int_{0}^{\delta}
|{\widehat \psi}_{0}|^{2}(v')dv' < {2 \over r_{0}}
\label{(2.13)}
\end{equation}
is fulfilled.
Such a condition requires an upper bound of the form \cite{KR2012}
\begin{equation}
|{\widehat \psi}_{0}| \leq {1 \over \sqrt{\delta}}.
\label{(2.14)}
\end{equation}
In order to control the error term $F$ in (2.4), we need for some positive $c$ \cite{KR2012}
\begin{equation}
{\rm tr}{\widetilde \chi}+{2 \over r}={\rm O}(\delta^{c}),
\; {dr \over du}+1={\rm O}(\delta^{c}), 
\; \eta={\rm O} \Bigr(\delta^{-{1 \over 2}}+c \Bigr),
\; \omega={\rm O}(\delta^{-1+c}),
\; \nabla \eta={\rm O} \Bigr(\delta^{-{1 \over 2}}+c \Bigr).
\label{(2.15)}
\end{equation}
Many optical structure equations (see Sect. 3) have curvature components as sources, and hence
one has to derive bounds not just for all Ricci coefficients $\chi,\omega,\eta,{\widetilde \chi},
{\widetilde \omega}$ and ${\widetilde \eta}$, but also for all null curvature components
$\alpha,\beta,\rho,\sigma,{\widetilde \alpha}$ and ${\widetilde \beta}$. In Ref. \cite{BH2009}
Christodoulou obtained such estimates by making the {\it short-pulse} ansatz for the initial 
data. This means that initial data are taken to be trivial and that 
${\widehat \psi}_{0}$ satisfies, relative to coordinates $v$ and transported coordinates $\omega$
along $H_{0}$ (transport being taken with respect to ${d \over dv}$), the condition
\begin{equation}
{\widehat \psi}_{0}(v,\omega)={1 \over \sqrt{\delta}}f_{0}(\delta^{-1}v,\omega),
\label{(2.16)}
\end{equation}
where $f_{0}$ denotes a fixed traceless, symmetric $S$-tangent $2$-tensor along $H_{0}$.

\section{The optical structure equations}
\setcounter{equation}{0}

The aim of this section is to summarize the conceptual and technical framework leading to
the $16$ optical structure equations of vacuum Einstein equations, since their knowledge is
not widespread, and the notation used in the literature is sometimes a bit cumbersome, so that
its potentialities are hidden rather than being fully appreciated.

We consider a spacetime manifold $(M,g)$ with boundary, where the metric
$g=\sum_{\mu,\nu=1}^{4}g_{\mu \nu} dx^{\mu} \otimes dx^{\nu}$ is taken to be a smooth solution
of the vacuum Einstein equations
$$
R_{\mu \nu}-{1 \over 2}g_{\mu \nu}R=0 \Longrightarrow R_{\mu \nu}=0.
$$
The past boundary of $M$ is the future null geodesic cone $C_{p}$ of a point $p$, and the
initial data are assigned on $C_{p}$. The future-directed null geodesics issuing from $p$ are the
generators of $C_{p}$ \cite{YF1952}, while a timelike geodesic from $p$ with tangent vector $T$ 
at $p$ is denoted by $\Gamma_{p}$. With the notation of the Introduction, let us define the vector
fields $Z \equiv \Omega e_{4}$ and $W \equiv \Omega e_{3}$. If $\xi$ is a $1$-form on
${M \over \Gamma_{p}}$ such that
\begin{equation}
\xi(Z)=\langle \sum_{m=1}^{2}\xi_{m}dx^{m}|\sum_{n=1}^{2}Z^{n}{\partial \over \partial x^{n}}\rangle
=\sum_{m=1}^{2}\xi_{m}Z^{m}=\xi(W)=\sum_{n=1}^{2}\xi_{n}W^{n}=0,
\label{(3.1)}
\end{equation}
we then say that $\xi$ is a $S$ $1$-form, which is therefore the specification of a $1$-form 
intrinsic to $S_{uv}$ for each $(u,v)$. The Lie derivative of $\xi$ with respect to $Z$ can be 
restricted to the tangent space $T S_{uv}$, and such a restriction is here denoted by
$\left . {\cal L}_{Z}\xi \right|_{T S_{uv}}$. This is a $S$ $1$-form as well as $\xi$.
Related geometrical concepts are as follows.
\vskip 0.3cm
\noindent
(i) A $S$ vector field is a vector field $X$ defined on ${M \over \Gamma_{p}}$ such that, at each
point $x \in {M \over \Gamma_{p}}$, $X$ is tangential to the surface 
$S_{uv}$ through $x$. This is therefore
a vector field intrinsic to $S_{uv}$ for each $(u,v)$.
\vskip 0.3cm
\noindent
(ii) A type $T_{s}^{q}$ $S$ tensor field $\theta$ is a type $T_{s}^{q}$ tensor field defined
on ${M \over \Gamma_{p}}$ such that, at each $x \in {M \over \Gamma_{p}}$ and each 
$X_{1},...,X_{s} \in T_{x}M$, one has
$$
\theta(X_{1},...,X_{s}) \in \otimes^{q}T_{x}S_{uv},
$$
and $\theta(X_{1},...,X_{s})=0$ if one of $X_{1},...,X_{s}$ is either $Z$ or $W$. One therefore deals
with a type $T_{s}^{q}$ tensor field intrinsic to $S_{uv}$ for each $(u,v)$.

The work in Ref. \cite{BH2009} proves that, for any given $S$ vector field $Y$, the Lie derivatives
$$
{\cal L}_{Z}Y=[Z,Y] \; {\rm and} \; {\cal L}_{W}Y=[W,Y]
$$
are also $S$ vector fields. One can therefore define the restricted Lie derivatives
\begin{equation}
\left . {\cal L}_{Z}Y \right|_{T S_{uv}} \equiv {\cal L}_{Z}Y, \;
\left . {\cal L}_{W}Y \right|_{T S_{uv}} \equiv {\cal L}_{W}Y. 
\label{(3.2)}
\end{equation}
As a next step, for a $S$ tensor field $\theta$ of type $T_{s}^{q}$, the Lie derivative
$\left . {\cal L}_{Z}\theta \right |_{T S_{uv}}$ is defined by considering $\theta$ on each
$H_{u}$ extended to the tangent space $TH_{u}$ according to the condition that it vanishes
if one of the entries is $Z$, and setting the Lie derivative of $\theta$ with respect to
$Z$, when restricted to the tangent space $T S_{uv}$, equal to the restriction to such a tangent 
space of the usual Lie derivative with respect to $Z$ of this extension. In analogous fashion, 
the restriction to the tangent space $T S_{uv}$ of the Lie derivative of $\theta$ with respect
to $W$ is defined by considering $\theta$ on each $H_{v}$ extended to the tangent space 
$T H_{v}$ in such a way that it vanishes if one of the entries is $W$, and setting
$$
\left . {\cal L}_{W}\theta \right |_{TS_{uv}}={\rm restriction} \; {\rm to} \; TS_{uv} 
\; {\rm of} \; {\rm the} \; {\rm usual} \; {\rm Lie} \; {\rm derivative} \; 
{\rm with} \; {\rm respect} \; {\rm to} \; W \; {\rm of} \; {\rm this} \;
{\rm extension}.
$$
This method yields Lie derivatives which are, themselves, $S$ vector fields of type
$T_{s}^{q}$. We write hereafter
\begin{equation}
\left . {\cal L}_{Z}\theta \right |_{TS_{uv}} \equiv D \theta, \;
\left . {\cal L}_{W}\theta \right |_{TS_{uv}} \equiv {\widetilde D} \theta.
\label{(3.3)}
\end{equation}
In particular, if $\theta$ is a $0$-form, i.e. a function $f$, one has
\begin{equation}
Df=Zf=\sum_{a=1}^{2}Z^{a}{\partial f \over \partial x^{a}}, \;
{\widetilde D}f=Wf=\sum_{b=1}^{2}W^{b}{\partial f \over \partial x^{b}}.
\label{(3.4)}
\end{equation}
For any function $f$ defined on ${M \over \Gamma_{p}}$, we denote by $d_{uv}f$ the $S$
$1$-form obtained by restriction to each surface $S_{uv}$ of the differential
$df$, i.e.
\begin{equation}
d_{uv}f \equiv \left . df \right |_{S_{uv}}
=\sum_{a=1}^{2}{\partial f \over \partial x^{a}}dx^{a}.
\label{(3.5)}
\end{equation}
This operation commutes with the $D$ and ${\widetilde D}$ derivatives, i.e. \cite{BH2009}
$$
D d_{uv}f=d_{uv}Df, \; {\widetilde D}d_{uv}f=d_{uv}{\widetilde D}f.
$$

Set now (cf. Sect. 1) ${\widehat L} \equiv e_{4} =\Omega L$,
${\widehat {\underline L}} \equiv e_{3}=\Omega M$. The tangent hyperplane $T_{p}H_{u}$ to a
given null hypersurface $H_{u}$ at a point $p \in H_{u}$ is given by all vectors $X$ at $p$
which are orthogonal to ${\widehat L}_{p}$, i.e.
\begin{equation}
T_{p}H_{u} \equiv \left \{ X \in T_{p}M: g(X,{\widehat L}_{p})=0 \right \},
\label{(3.6)}
\end{equation}
while the tangent hyperplane $T_{p}H_{v}$ is given by
\begin{equation}
T_{p}H_{v} \equiv \left \{ X \in T_{p}M: g(X,{\widehat {\underline L}}_{p})=0 \right \}.
\label{(3.7)}
\end{equation}
Since $H_{u}$ and $H_{v}$ are null hypersurfaces in spacetime, their induced metrics are degenerate,
while the induced metric $h$ on each surface $S_{uv}$ is Riemannian (i.e. positive-definite), and
is a symmetric 2-covariant tensor field
$$
h=\sum_{a,b=1}^{2}h_{ab}dx^{a} \otimes dx^{b}, \;
h_{ab}=h_{(ab)}.
$$
Any vector $X \in T_{p}H_{u}$ can be uniquely decomposed into a vector collinear to
${\widehat L}_{p}$ and a vector tangent to the surface $S_{uv}$, i.e. ($a$ being a real number)
\begin{equation}
X \in T_{p}H_{u} \Longrightarrow X=a {\widehat L}_{p}+PX, \; PX \in T_{p} S_{uv}.
\label{(3.8)}
\end{equation}
If $X$ and $Y$ are any two vectors tangent to $H_{u}$ at $p$, one has a simple relation 
between spacetime metric $g$ and induced metric $h$, i.e.
\begin{equation}
g(X,Y)=h(PX,PY).
\label{(3.9)}
\end{equation}
Similarly, one has
\begin{equation}
X \in T_{p}H_{v} \Longrightarrow X=a {\widehat {\underline L}}_{p}+\pi X, 
\; \pi X \in T_{p} S_{uv},
\label{(3.10)}
\end{equation}
and
\begin{equation}
g(X,Y)=h(\pi X, \pi Y)
\label{(3.11)}
\end{equation}
for any pair of vectors $X$ and $Y$ tangent to $H_{v}$ at $p$.

The second fundamental form $\chi$ of a null hypersurface $H_{u}$ is a bilinear form
$$
\chi_{u}: T_{p}H_{u} \times T_{p}H_{u} \rightarrow {\bf R}
$$
defined by (see components in (1.15) and (1.19))
\begin{equation}
\chi_{u}(X,Y) \equiv g(\nabla_{X}{\widehat L},Y).
\label{(3.12)}
\end{equation}
It can be shown to be symmetric, because \cite{BH2009}
\begin{equation}
\chi_{u}(X,Y)-\chi_{u}(Y,X)=-g({\widehat L},[X,Y])=0,
\label{(3.13)}
\end{equation}
where $X$ and $Y$ are extended to vector fields along $H_{u}$ which are
tangential to $H_{u}$. It should be stressed that $\chi$ is intrinsic to $H_{u}$, because the
vector field ${\widehat L}$ is tangential to $H_{u}$. One has
\begin{equation}
\chi_{u}(X,Y)=\chi_{u}(PX,PY),
\label{(3.14)}
\end{equation}
and hence $\chi_{u}$ is a symmetric $2$-covariant $S$ tensor field. Similarly, for the null
hypersurfaces $H_{v}$ one defines
$$
\chi_{v}: T_{p}H_{v} \times T_{p}H_{v} \rightarrow {\bf R}
$$
such that
\begin{equation}
\chi_{v}(X,Y) \equiv {\widetilde \chi}(X,Y)  
\equiv g(\nabla_{X}{\widehat {\underline L}},Y)
=\chi_{v}(Y,X)=\chi_{v}(\pi X,\pi Y).
\label{(3.15)}
\end{equation}

If $\theta$ is any $2$-covariant $S$ tensor field, we denote by $\theta^{\sharp}$ the $S$ tensor
field of type $(1,1)$ (i.e. once covariant and once contravariant) such that
\begin{equation}
h(\theta^{\sharp}X,Y)=\theta(X,Y) \; \; 
\forall X,Y \in T_{p}S_{uv}.
\label{(3.16)}
\end{equation}
If $e_{1},e_{2}$ is an arbitrary basis for the tangent space $T_{p}S_{uv}$, one
has (unlike Ref. \cite{BH2009}, we do not use block capital letters for tensor components
here, so as to avoid confusion with two-component spinors for which $A,B$ are a standard notation
for unprimed spinor indices \cite{PE1965})
\begin{equation}
\theta^{\sharp}e_{a}=\sum_{b=1}^{2}(\theta^{\sharp})_{a}^{b} \; e_{b}, \;
(\theta^{\sharp})_{a}^{b}=\sum_{c=1}^{2}\theta_{ac}(h^{-1})^{cb}, 
\; a,b=1,2,
\label{(3.17)}
\end{equation}
which verify indeed the explicit form of Eq. (3.16), i.e.
\begin{equation}
\sum_{a,b,c=1}^{2}h_{ab}(\theta^{\sharp})_{c}^{a} \; X^{c}Y^{b}
=\sum_{a,b=1}^{2}\theta_{ab}X^{a}Y^{b}.
\label{(3.18)}
\end{equation}

If $\theta$ and $T$ are symmetric $2$-covariant $S$ tensor fields, their product 
$\theta \times T$ is defined by
\begin{equation}
(\theta \times T)(X,Y) \equiv h(\theta^{\sharp}X,T^{\sharp}Y), \; \;
\forall X,Y \in T_{p}S_{uv}.
\label{(3.19)}
\end{equation}
In arbitrary local coordinates for $S_{uv}$, this formula reads as
\begin{equation}
(\theta \times T)_{ab}=\sum_{r,s=1}^{2}h_{rs}
(\theta^{\sharp})_{a}^{r} (T^{\sharp})_{b}^{s}
=\sum_{r,s=1}^{2}(h^{-1})^{rs}\theta_{ar}T_{bs}.
\label{(3.20)}
\end{equation}

The optical structure equations will involve (see below) the rescaled tensor fields 
(see (1.15) and (1.19)-(1.21))
\begin{equation}
\chi_{ab}' \equiv \Omega^{-1}\chi_{ab}, \;
{\widetilde \chi}_{ab}' \equiv \Omega^{-1} {\widetilde \chi}_{ab},
\label{(3.21)}
\end{equation}
and a hypersurface version of divergence, curl and covariant derivative. More
precisely, one defines the covariant derivative intrinsic to $S_{uv}$,
for any pair $X,Y$ of $S$ vector fields, with the help of a projection 
operator $\pi$ to the surfaces $S_{uv}$, as given by \cite{BH2009}
\begin{equation}
\left({ }_{uv}\nabla \right)_{X}Y=\pi \nabla_{X}Y, \;
\pi V=V+{1 \over 2}g(V,e_{3})e_{4}+{1 \over 2}g(V,e_{4})e_{3}
\; \in T_{q}S_{uv},
\label{(3.22)}
\end{equation}
for all $V \in T_{q}\left({M \over \Gamma_{p}}\right)$. 
Furthermore, the intrinsic divergence is defined by the formula
\begin{equation}
\left({ }_{uv}{\rm div}\theta \right)_{a} \equiv \sum_{c=1}^{2}\left({ }_{uv}\nabla\right)_{c}
\theta_{\; a}^{\sharp c},
\label{(3.23)}
\end{equation}
and one denotes by ${ }_{uv}\varepsilon$ the area $2$-form of $S_{uv}$, with components
\begin{equation}
\left({ }_{uv}\varepsilon \right)_{ab}
=\left({ }_{uv}\varepsilon \right)(e_{a},e_{b}), \; a,b=1,2.
\label{(3.24)}
\end{equation}
The latter concept is used to define the intrinsic curl of a $S$ $1$-form $\xi$ according to
\begin{equation}
{ }_{uv}{\rm curl}\xi \equiv {1 \over 2} \sum_{a,b,c,d=1}^{2}
\left({ }_{uv}\varepsilon \right)_{cd}(h^{-1})^{ac}
(h^{-1})^{bd}\left({ }_{uv}\nabla_{a}\xi_{b}
-{ }_{uv}\nabla_{b}\xi_{a}\right).
\label{(3.25)}
\end{equation}
Out of the area $2$-form of $S_{uv}$ one can also build the twice sharp $\varepsilon$,
defined as
\begin{equation}
\left({ }_{uv}\varepsilon^{\sharp \sharp}\right)^{ab} \equiv 
\sum_{c,d=1}^{2}\left({ }_{uv}\varepsilon \right)_{cd}
(h^{-1})^{ac} (h^{-1})^{bd},
\label{(3.26)}
\end{equation}
and hence the wedge product of symmetric $2$-covariant $S$ tensor fields $\theta$ and $T$,
i.e. \cite{BH2009}
\begin{equation}
\theta \wedge T \equiv \sum_{a,b,c,d=1}^{2}
\left({ }_{uv}\varepsilon^{\sharp \sharp}\right)^{ab}
(h^{-1})^{cd}\theta_{ac} T_{bd}.
\label{(3.27)}
\end{equation}
Last, one considers
\begin{equation}
(\xi,\xi') \equiv \sum_{a,b=1}^{2}(h^{-1})^{ab}\xi_{a}\xi_{b}',
\label{(3.28)}
\end{equation}
\begin{equation}
|\xi| \equiv \sqrt{(\xi,\xi)}.
\label{(3.29)}
\end{equation}
 
Since we have defined all concepts that are needed, 
we can now write down from Ref. \cite{BH2009}, but with
our notation, the $16$ equations expressing the optical structure of vacuum Einstein equations.
They read as follows:
\begin{equation}
Dh=2 \Omega \chi, \; {\widetilde D}h=2 \Omega {\widetilde \chi},
\label{(3.30)}
\end{equation}
\begin{equation}
D \chi'=\Omega^{2} \chi' \times \chi' -\alpha ,
\label{(3.31)}
\end{equation}
\begin{equation}
{\widetilde D}{\widetilde \chi}'=\Omega^{2} {\widetilde \chi}' \times {\widetilde \chi}'
-{\widetilde \alpha},
\label{(3.32)}
\end{equation}
\begin{equation}
D \eta=\Omega \left(\chi^{\sharp} \cdot {\widetilde \eta}-\beta \right),
\label{(3.33)}
\end{equation}
\begin{equation}
{\widetilde D} {\widetilde \eta}=\Omega \left({\widetilde \chi}^{\sharp} \cdot \eta
+{\widetilde \beta} \right),
\label{(3.34)}
\end{equation}
\begin{equation}
D {\widetilde \omega}=\Omega^{2}\Bigr[2 (\eta,{\widetilde \eta})
-|\eta|^{2}-\rho \Bigr],
\label{(3.35)}
\end{equation}
\begin{equation}
{\widetilde D} \omega=\Omega^{2}\Bigr[2 (\eta,{\widetilde \eta})
-|\eta|^{2}-\rho \Bigr],
\label{(3.36)}
\end{equation}
\begin{equation}
K+{1 \over 2}({\rm tr}\chi)({\rm tr}{\widetilde \chi})
-{1 \over 2}(\chi,{\widetilde \chi})=-\rho ,
\label{(3.37)}
\end{equation}
\begin{equation}
{ }_{uv}{\rm div}\chi'-{ }_{uv}d({\rm tr}\chi')
+{\chi'}^{\sharp}\cdot \eta-({\rm tr}\chi')\eta=-\Omega^{-1}\beta ,
\label{(3.38)}
\end{equation}
\begin{equation}
{ }_{uv}{\rm div}{\widetilde \chi}'-{ }_{uv}d({\rm tr}{\widetilde \chi}')
+{\widetilde {\chi'}}^{\sharp}\cdot {\widetilde \eta}
-({\rm tr}{\widetilde \chi}'){\widetilde \eta}=\Omega^{-1}{\widetilde \beta} ,
\label{(3.39)}
\end{equation}
\begin{equation}
{ }_{uv}{\rm curl}{\widetilde \eta}={1 \over 2}\chi \wedge {\widetilde \chi} -\sigma ,
\label{(3.40)}
\end{equation}
\begin{equation}
{ }_{uv}{\rm curl}\eta={ }_{uv}{\rm curl}\zeta=-{ }_{uv}{\rm curl}{\widetilde \eta},
\label{(3.41)}
\end{equation}
\begin{equation}
D(\Omega {\widetilde \chi})=\Omega^{2}\Bigr[{ }_{uv}\nabla{\widetilde \eta}
+{ }_{uv}{\widetilde \nabla}{\widetilde \eta}
+2 {\widetilde \eta} \otimes {\widetilde \eta}
+{1 \over 2}(\chi \times {\widetilde \chi}+{\widetilde \chi} \times \chi)
+\rho h \Bigr],
\label{(3.42)}
\end{equation}
\begin{equation}
{\widetilde D}(\Omega \chi)=\Omega^{2}\Bigr[{ }_{uv}\nabla \eta
+{ }_{uv}{\widetilde \nabla} \eta
+2 \eta \otimes \eta
+{1 \over 2}(\chi \times {\widetilde \chi}+ \chi \times \chi)
+\rho h \Bigr],
\label{(3.43)}
\end{equation}
\begin{equation}
{\widetilde D}\eta=-\Omega \Bigr({\widetilde \chi}^{\sharp} \cdot \eta
+{\widetilde \beta}\Bigr)+2 d_{uv} {\widetilde \omega},
\label{(3.44)}
\end{equation}
\begin{equation}
D{\widetilde \eta}=-\Omega \Bigr(\chi^{\sharp} \cdot {\widetilde \eta}
-\beta\Bigr)+2 d_{uv} \omega.
\label{(3.45)}
\end{equation}
The first of Eqs. (3.30), and Eqs. (3.31), (3.33), (3.35), are propagation equations along the
generators of each $H_{u}$; the second of Eqs. (3.30), and Eqs. (3.32), (3.34), (3.36), are
propagation equations along the generators of each $H_{v}$. Moreover, Eq. (3.37) is the Gauss
equation of the embedding of the surfaces $S_{uv}$, with Gauss curvature $K$, in the spacetime
manifold $(M,g)$, while Eqs. (3.38) and (3.39) are the Codazzi equations of such an embedding. 

\section{Open problems}

As far as we can see, at least two outstanding problems deserve careful consideration:
\vskip 0.3cm
\noindent 
(i) Despite the elegant proof obtained in Refs. [3] and [4] that the non-linear equations
leading to trapped surface formation can be studied by discarding some terms, it would
be interesting to solve them numerically without discarding any term, no matter how small it can be.
\vskip 0.3cm
\noindent
(ii) It would be interesting to apply optical-structure methods to extended theories of 
classical gravity, and possibly to the quantum theory of black holes.

\section*{Acknowledgments}

G. E. is grateful to the Dipartimento di Fisica ``Ettore Pancini'' of Federico II University for
hospitality and support.

\end{document}